\documentclass[aps,prd,twocolumn,superscriptaddress,nofootinbib,showpacs]{revtex4-2}
\usepackage[pass,letterpaper]{geometry}
\pdfoutput=1 
\usepackage[]{graphicx}
\usepackage{color}
\usepackage{slashed}
\usepackage{latexsym,amsmath,amsfonts,amssymb,bm}
\usepackage[pdfencoding=auto]{hyperref}
\usepackage{cancel}
\usepackage{booktabs}
\usepackage{verbatim}
\usepackage{hyperref}

\def\dd{\textrm{d}}

\begin{document}

\title{Hubble tension as a window on the gravitation of the dark matter sector}  

\author{Cyril Pitrou}
\email{pitrou@iap.fr}
\affiliation{Institut d'Astrophysique de Paris, CNRS UMR 7095,
Sorbonne Universit\'e, 98 bis Bd Arago, 75014 Paris, France}

\author{Jean-Philippe Uzan}
\email{uzan@iap.fr}
\affiliation{Institut d'Astrophysique de Paris, CNRS UMR 7095,
Sorbonne Universit\'e, 98 bis Bd Arago, 75014 Paris, France}

\date{\today}  
\begin{abstract}
A simple and minimal extension of the standard cosmological $\Lambda$CDM model in which dark matter experiences an additional long-range scalar interaction is demonstrated to alleviate the long-lasting Hubble-tension while primordial nucleosynthesis predictions remain unaffected and passing by construction all current local tests of general relativity. The theoretical formulation of this $\Lambda\beta$CDM model and its comparison to astrophysical observations are presented to prove its ability to fit existing data and potentially resolve the tension.
\end{abstract}

\maketitle  
\noindent{\bf Introduction.} The Hubble tension questions the status of the standard $\Lambda$CDM cosmological model. It arises from the discrepancy between  the model-dependent determination of  $H_0$ from the {\em Planck} analysis of the cosmic microwave background (CMB)  combined with baryon acoustic oscillations (BAO) and the Hubble diagram interpretation, in particular from the SH0ES experiment~\cite{Riess:2016jrr}, that is almost independent of physical assumptions. The former leads to the value  $H_0=(67.49 \pm0.53)$~km/s/Mpc~\cite{Planck:2018vyg} while the latter concludes that $H_0 = (73.04 \pm1.04)$~km/s/Mpc~\cite{Riess:2021jrx}.  This results in a $\simeq4.8\sigma$ tension on $H_0$. This letter investigates a new road by focusing on the properties of gravitation in the dark matter (DM) sector and proposing a theory that offers a minimal extension of the $\Lambda$CDM that avoids {\em by construction} the existing constraints.

\vskip0.25cm
\noindent{\bf The Hubble tension.} Consider a Friedman-Lema\^{\i}tre spacetime with metric, $\dd s^2 = -\dd t^2 + a^2(t)\gamma_{ij}\dd x^i \dd x^j$ where $\gamma_{ij}$ is the spatial metric and $a$ the scale factor. With $a_0=1$, the redshift and Hubble function are $1+z=1/a$ and $H=(\ln a)^.$, a dot referring to a derivative with respect to the cosmic time. The $H_0$ problem is often formulated~\cite{Schoneberg:2021qvd} as a low/high-redshift tension. Indeed, in first approximation, the key physical parameters  are the (comoving) sound horizon
\begin{equation}\label{e.rs}
r_s=\frac{1}{H_0}\int_{z_*}^\infty c_s E^{-1/2}(z) \dd z 
\end{equation}
where $E(z)\equiv H/H_0$ and $z_*\sim 1088$ at recombination, and the comoving angular diameter distance,
\begin{equation}\label{e.rs+1}
  R_{\rm ang}=\frac{1}{H_0}f_K\left[\int_0^z E^{-1/2}(z) \dd z \right]\,.
\end{equation}
Since their ratio fixes the physical angular scales of the acoustic peaks, most of the arguments on the $H_0$ tension circle around the sound horizon with two main categories of models. ``Late time solutions" modify the expansion history after recombination, increasing $H_0$ while keeping $r_s$ unchanged, while ``early time solutions" modify it before recombination changing both $H_0$ and $r_s$. While their relative statistical merits have been compared~\cite{DiValentino:2021izs,Schoneberg:2021qvd} it has been pointed out~\cite{Jedamzik:2020zmd}  that models reducing $r_s$ can never fully resolve the Hubble tension, if they are expected to also be in agreement with other cosmological datasets. The minimal model presented in this {\em letter} will, as we shall demonstrate, keep $r_s$ identical to its $\Lambda$CDM value but with a higher $H_0$ at the expense of a lower $\Omega_{\rm D0}$.  Hence we present a model that (1) does not change standard physics, and (2) keeps the CMB physics unchanged.  This points toward a modification of the physics of the DM sector around recombination that would play on $(H_0,\Omega_{\rm D0})$.

\vskip0.25cm
\noindent{\bf Towards a minimal extension.}  We set ourselves the constraints that the new theory should have no effect on primordial nucleosynthesis (BBN) and in all tests of general relativity (GR) including violation of the weak equivalence principle (WEP)~\cite{MICROSCOPE:2022doy} and variation of the constants~\cite{Uzan:2002vq}. This implies that we need to avoid any new interaction in the visible sector of the Standard model (SM) and that any new degree of freedom shall have a negligible energy density so that it does not directly affect the expansion history. While the window is small, we still have the possibility to introduce a DM ``fifth force''.

We assume that DM  enjoys a scalar-tensor (ST) theory while the SM sector is subjected to GR. This can be seen as a subclass of models in which a light dilaton couples nonuniversally to the SM and DM fields but those are strongly constrained~\cite{Damour:1990tw,Coc:2008yu,Fuzfa:2007sv} by  BBN and that our model evades {\em by construction}. Indeed the DM sector will witness a  time variation of its gravitational constant but it cannot be measured directly and does not affect BBN during which DM is subdominant. The DM density in our local environment is estimated~\cite{deSalas:2020hbh} to be between $0.4$ and $0.6$~GeV/cm$^3$, too small a value to have observable dynamical effects. To finish, the change in the strength of gravity in the DM sector shall alleviate the $H_0$ tension. It is clear that the new field will have fluctuations so that we need to treat its background and perturbation effects to consistently predict its cosmological effects.

\vskip0.25cm
 \noindent{\bf Definition.}  The theory for this  minimal and simple extension of the  $\Lambda$CDM  is described by the action $S = S_{\rm GR} + S_{\rm SM} + S_{\varphi} + S_{\rm D}$ with a  new light scalar degree of freedom\footnote{We use the normalisation of ST theories~\cite{Damour:1992we}.} $\varphi$ mediating a long-range interaction for DM. The actions for the visible sector are
\begin{equation}
 S_{\rm GR} +S_{\rm SM}=  \int \dd^4x\sqrt{-g}\left[\frac{R - 2\Lambda}{16\pi G}+{\cal L}_{\rm SM}[\psi; g_{\mu\nu}] \right] \label{e.LRG}
\end{equation}
while the DM sector is modeled by
 \begin{eqnarray}
 S_{\varphi}  &=&- \int \frac{\dd^4x}{16\pi G}\sqrt{-g}\left[  2g^{\mu\nu} \partial_\mu\varphi\partial_\nu\varphi +  4V(\varphi)\right] \\
 S_{\rm D} &=& \int \dd^4x \sqrt{-\tilde g} {\cal L}_{\rm D}[\psi; \tilde g_{\mu\nu}],\label{TH-5}
\end{eqnarray}
with the DM-metric $\tilde g_{\mu\nu}=A^2(\varphi)g_{\mu\nu}$. ${\cal L}_{\rm SM/D}$ are the Lagrangians of the SM, DM sectors. The free functions $V$ and $A$ are the field potential and coupling to DM. This theory is defined in the {\it SM frame}  in which nothing departs from GR for the SM with which all tests have been performed so far. The equations of motion are fully described in Ref.~\cite{PU_2}. They are identical to GR, except for the DM sector for which
\begin{eqnarray}
&& \nabla_\mu T_{\rm(DM)}^{\mu\nu} =  \alpha(\varphi) T^{\rm(DM)}_{\sigma\rho} g^{\sigma\rho} \partial^\nu\varphi\label{FEq_3}  \\
&& \Box \varphi = \frac{\dd V}{\dd\varphi} - \frac{\kappa}{2} \alpha(\varphi) T^{\rm(DM)}_{\mu\nu} g^{\mu\nu}\label{FEq_4}
\end{eqnarray}
where we have defined $\alpha(\varphi) \equiv {\dd \ln A}/{\dd\varphi}$.

\vskip0.25cm
 \noindent{\bf Cosmological dynamics.} At the background level, the Einstein equations yield the Friedmann equation
\begin{eqnarray}
 && 3\left(H^2 + \frac{K}{a^2}\right) =    \kappa( \rho +   \rho_{\rm D}+\rho_\varphi)+ \Lambda \label{einsteinEF1}
\end{eqnarray}
with $\kappa \rho_{\varphi} = \dot\phi^2+2V$ and $\kappa\equiv 8\pi G$. While the conservation equations in the SM sector remain unchanged, for DM they become
\begin{equation}\label{e.c2}
\dot\rho_{\rm D} + 3H \rho_{\rm D}= \alpha (\varphi) \rho_{\rm D}\dot \varphi,\quad
 \ddot\varphi + 3H\dot\varphi = -\frac{\dd V}{\dd\varphi} -\frac{\kappa}{2} \alpha  \rho_{\rm D},
 \end{equation}
which gives $\rho_{\rm D} \propto a^{-3} A$ so that 
\begin{equation}\label{e.rhoDM}
 G\rho_{\rm D} =  G\rho_{\rm D0} a^{-3}\left[ 1+ \delta_A(\varphi)\right] \equiv G_{\rm eff}\rho_{\rm D0} a^{-3}
\end{equation}
with $\delta_A\equiv \left(A/A_0 -1\right)$. Hence $\varphi$ triggers a dynamical effective gravitational constant $G_{\rm eff}$. However, it does not correspond to what would be defined as the gravitational constant in e.g. a Cavendish experiment~\cite{Damour:1992we}. As usual, we define the cosmological fractions $\Omega_{i 0}={8\pi G \rho_{i 0}}/{3H_0^2}$ for the baryons (b), the radiation (r), DM and $\Omega_{\Lambda 0}={\Lambda}/{3H_0^2}$ while $\Omega_{K 0}=-{K}/{3H_0^2}$ is assumed to be $0$,  so that 
\begin{eqnarray}
&&E^2(z) = (\Omega_{\rm b0}+\Omega_{\rm D0} )(1+z)^3 + \Omega_{\rm r0}(1+z)^4 + \Omega_{\Lambda0} \nonumber \\
&& \qquad\qquad + \Omega_{\rm D0}(1+z)^3\delta_A (z)+ \Omega_{\dot\varphi} + \Omega_V.
\end{eqnarray}
with $\Omega_{\dot\varphi} ={\dot\varphi^2}/{3H_0^2}$ and  $\Omega_V = {2V}/{3H_0^2}$. The first line corresponds to the standard $\Lambda$CDM while the second gathers all the effects of the scalar interaction.

\vskip0.25cm
{\bf Generic properties of the model.} For the sake of demonstrating the power of this model to fit the data, this letter is restricted to the minimal $\Lambda\beta$CDM that assumes a massless ($V=0$, hence $P_\varphi=\rho_\varphi$) scalar field with $A(\varphi)= 1+\frac{1}{2}\beta\varphi^2$.
\begin{figure}[htb]
 	\centering
 	\includegraphics[width=0.5\textwidth]{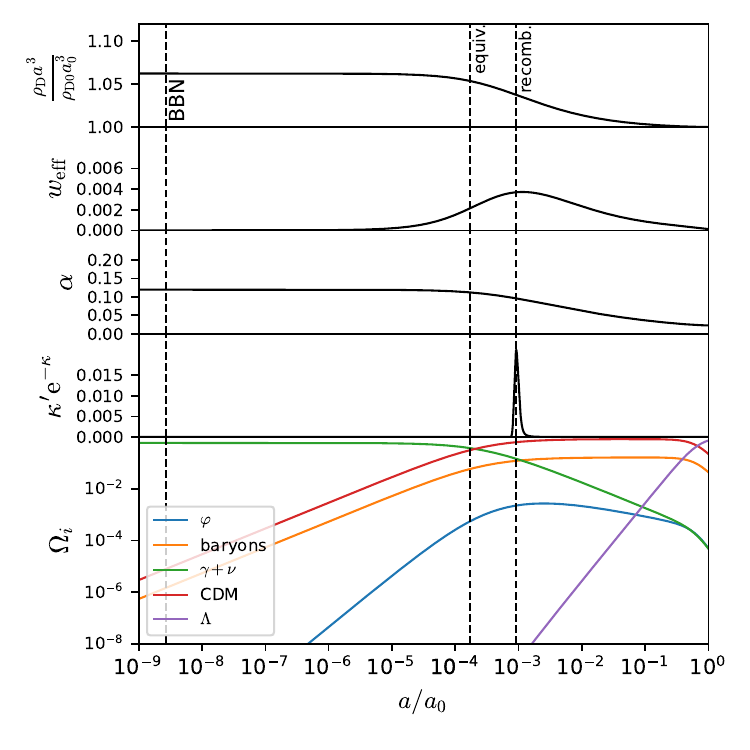}
		\vspace{-0.75cm}
 	\caption{From top to bottom: $\rho_{\rm D}$ which exhibits a  departure from pure dust between equivalence and recombination, the effective equation of state~(\ref{e.weff}),   the strength of the scalar coupling $\alpha$ that shows that DM gravity is stronger at high $z$ but similar as in the visible sector today, the CMB visibility function, and the evolution of the energy densities in units of $3H^2(a)/8\pi G$ proving that $\varphi$ remains subdominant at all times. All curves correspond to the best fit parameters of the base+BAO($z>1$)+$H_0$ dataset, i.e. $\Omega_{\rm b0}h^2= 0.02250$, $\Omega_{\rm m}=0.2619$, $h=0.7240$, $\ln(10^{10}A_s)=3.0569$, $n_s=0.9741$, $\tau_{\rm reio}=0.0601$, $\beta=0.2509$ and $\varphi_i=0.7159$.}
 	\label{fig:2}
 	\vspace{-0.3cm}
 \end{figure}
 Fig.~\ref{fig:2}  confirms that in $\Lambda\beta$CDM, $\rho_\varphi$ is negligible (it is at most $0.27\%$ of the total matter content around $z=z_*$), i.e. $\varphi$ modifies the strength of gravity for  DM but not the expansion history by its stress-energy. It is thus not a dark energy model. Then, the DM scalar force vanishes in the late universe ($\alpha\rightarrow0$) and saturates to $\alpha\simeq 0.12$ in the early universe, which corresponds to a change of the strength of gravity in the DM sector of $1.4\%$ while $G_{\rm eff}$ undergoes a $A_\infty/A_0-1\sim 6.4\%$ variation roughly between equivalence and recombination. Phenomenologically this can be described by an effective DM equation of state  from Eq.~(\ref{e.c2}) by $\dot\rho_{\rm D}+3H\rho_{\rm D} [1+w_{\rm eff}(a)]=0$,
\begin{equation}\label{e.weff}
 w_{\rm eff}(a) = -\frac{1}{3} \frac{\dd\ln A}{\dd\ln a}\,.
\end{equation}
Fig.~\ref{fig:2} shows that it departs from $0$ only between $z=10$ and $z=10^5$ making this minimal model similar to the standard $\Lambda$CDM when structures form. The model does not fall in the early/late categories; the scalar interaction being controlled by $\rho_{\rm D}$ naturally occurs shortly after equivalence and before recombination. This is a generic feature of our models. The conservation equations imply that it can be interpreted as if $\rho_{\rm D}$ transfers to $\rho_\varphi$ that redshifts faster as $a$ grows and even faster than radiation at small $z$.  This is a key difference with models of DM decaying into dark radiation that scales as $a^{-4}$ at all times that generally predict a suppression of the matter power spectrum~\cite{Schoneberg:2021qvd} avoided in our model.  To finish, the {\em key feature} of  $\Lambda\beta$CDM is that it keeps both $r_s$ and the distance to the last scattering surface (LSS) unchanged while having a higher $H_0$ at the cost of a lower $\Omega_{\rm D0}$ at low $z$, leading for the best fit to a younger universe of 13.51~Gyr instead of 13.79~Gyr. Since $\Omega_{\rm b0}/\Omega_{\rm r0}$ is fixed by BBN, the relative heights and shapes of the CMB peaks are almost unaffected, as confirmed in the residuals displayed in Fig.~\ref{fig:3}. 
 \begin{figure}[tb]
 	\centering
	 \includegraphics[width=0.5\textwidth]{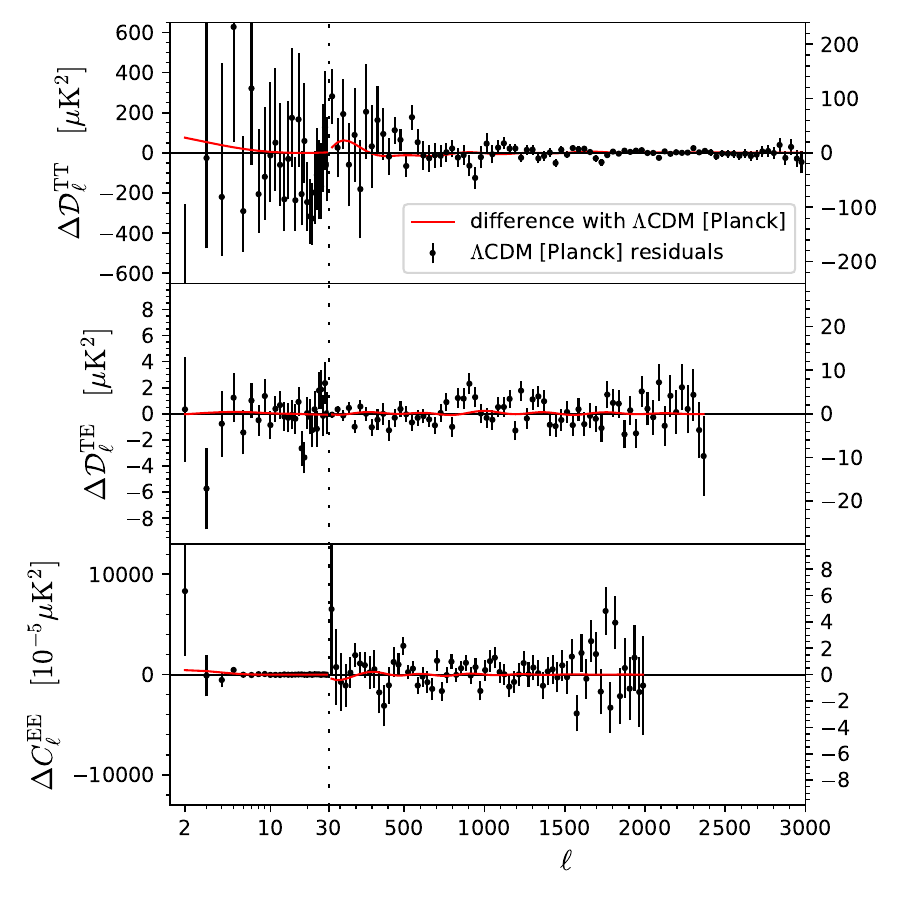}
		\vspace{-0.75cm}
 	\caption{{\it Black:} residuals of TT, TE and EE spectra of the $\Lambda$CDM best fit (from {\em Planck} data only)~\cite{Planck:2018vyg} with Planck data. {\it Red:} Differences between our $\Lambda\beta$CDM best fit spectra (obtained with the base dataset+BAO($z>1$)+$H_0$) and the former spectra.}
 	\label{fig:3}
 	\vspace{-0.25cm}
 \end{figure}

\vskip0.25cm 
 \noindent{\bf Comparison to data.}  The model is implemented both at the background and perturbations levels as a modification of the CLASS code \cite{CLASS} (see our companion article~\cite{PU_2} for details) in order to test its power on the $H_0$ tension through a MCMC analysis with Cobaya~\cite{Cobaya}. We consider a baseline dataset that consists in CMB data from {\em Planck} \cite{Planck:2019nip} (low and high $\ell$ temperature, polarization and lensing); weak lensing data from DES Y1 \cite{DES:2017myr}, and supernovae (SN) data from Pantheon~\cite{Pan-STARRS1:2017jku}. BAO data are subsequently added to this baseline in two different ways. First, we split BAO measurements into a low-$z$ dataset ($z<1$), which consists in 6dF \cite{2011MNRAS.416.3017B}, SDSS DR7 \cite{Ross:2014qpa}, DR12 \cite{BOSS:2016wmc} and the luminous red galaxies (LRG) of SDSS DR16 \cite{eBOSS:2020yzd}, and a high-$z$ dataset~\cite{eBOSS:2020yzd} from SDSS DR16 emission line galaxies (ELG), quasistellar objects (QSO) and Lyman-$\alpha$ (Ly$\alpha$) absorption lines. Whenever redshift space distorsions (RSD) are available, they are added to these datasets. Either BAO data from all redshifts is considered, or only BAO data from the high-$z$ set, noted BAO$(z>1)$. Finally we add a prior on $H_0$ from the latest SH0ES results~\cite{Riess:2021jrx}. All cosmological parameters are varied in an Euclidean cosmology, along with $\beta$ and the attractor initial value $\varphi_i$ of the scalar field.
  \begin{figure}[tb]
 	\centering
 	\includegraphics[width=0.5\textwidth]{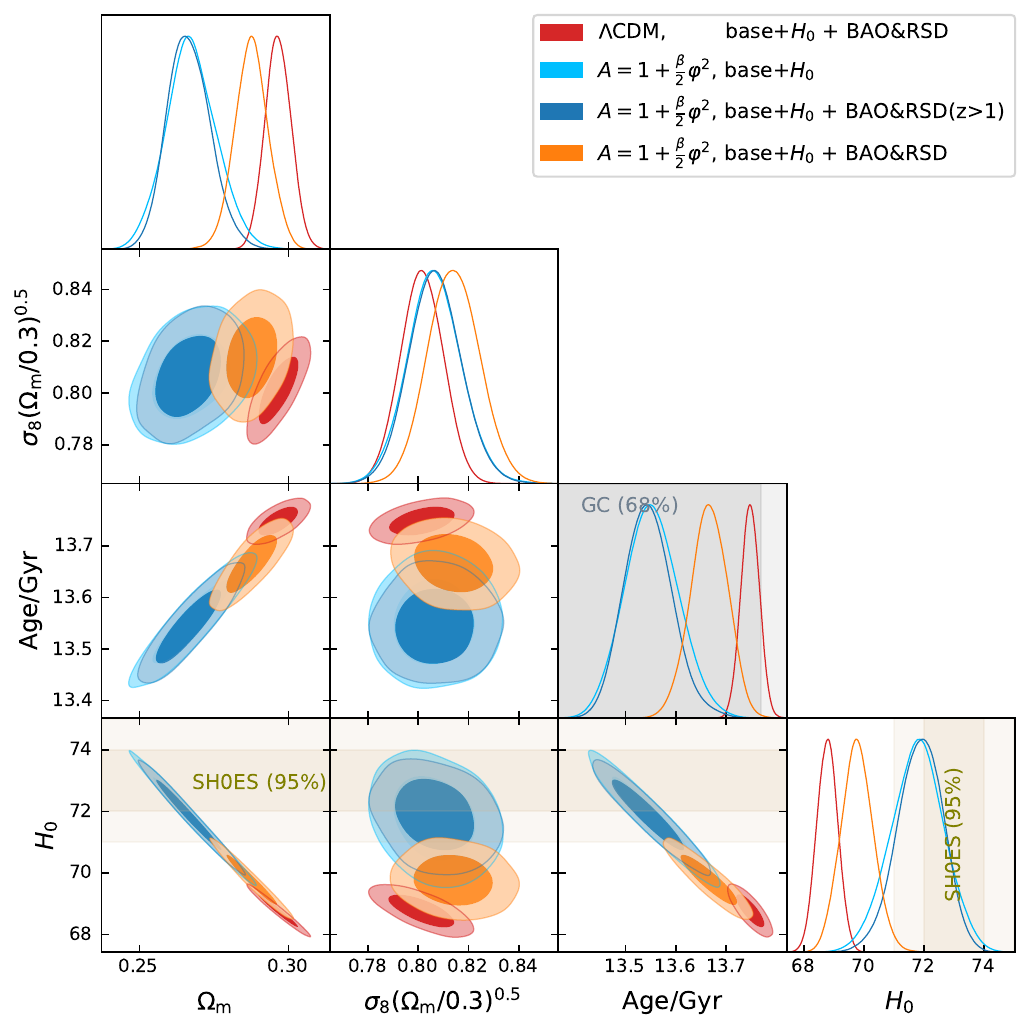}
	 	\vspace{-0.75cm}
 	\caption{Statistical comparison of the $\Lambda$CDM and $\Lambda\beta$CDM to cosmological data focusing on the four parameters $(H_0, \Omega_{\rm m},S_8,t_{\rm U}$) in order to highlight the Hubble tension. See Ref.~\cite{PU_2} for a full analysis and details. Plots were performed with GetDist~\cite{Lewis:2019xzd}. }
 	\label{fig:4}
 	\vspace{-0.25cm}
 \end{figure} 

\begin{table*}[!htp]
    
  \centering
    \resizebox{\textwidth}{!}{
\begin{tabular}{|ll|ccccc|ccc|}
\hline
Model & base+$H_0$+& $\Omega_{\rm m}$ & $\Omega_{\rm b0} h^2$ & $h$  & $S_8$   &  Age (Gyr) & $H_0$ tension  & $Q_{\rm DMAP}$ &$\Delta$AIC \\
\hline
$\Lambda$CDM & BAO &   $0.2965\pm0.0044$ & $0.02263\pm0.00013$   &  $0.6877\pm0.0035$ & $0.801\pm0.009$ & $13.75\pm0.02$  & 4.4$\sigma$& 4.8 & 0 \\
$\Lambda$CDM & BAO($z>1$) &   $0.2912\pm0.0052$ & $0.02270\pm0.00014$   &  $0.6919\pm0.0042$ & $0.794\pm0.010$ & $13.73\pm0.02$  & 4.1$\sigma$& 4.4 & 0 \\
$\Lambda\beta$CDM & BAO &   $0.2875\pm0.0056$ & $0.02249\pm0.00014$   &  $0.6977\pm0.0054$ & $0.814\pm0.010$ & $13.67\pm0.04$  & 3.8$\sigma$& 3.6 & -2.6 \\
$\Lambda\beta$CDM & BAO($z>1$) &   $0.2666\pm0.0073$ & $0.02246\pm0.00015$   &  $0.7187\pm0.0076$ & $0.807\pm0.010$ & $13.55\pm0.05$  & 1.8$\sigma$& 2.0 & -14.5  \\
\hline 
\end{tabular}}
\caption{Comparison of the posterior marginals and success criteria of the standard $\Lambda$CDM and $\Lambda\beta$CDM.
  The $H_0$ tension is the Gaussian tension evaluated between the marginal constraint from the model with the dataset without SH0ES, and the constraint from SH0ES alone. The $\Delta$AIC is the difference in maximum $\chi^2$ between the model and the $\Lambda$CDM with the considered dataset (hence including $H_0$ from SH0ES). \label{tab1}}
\end{table*}%
 
The results in the space $(H_0,\Omega_{\rm m}, S_8,t_{\rm U})$, where $\Omega_{\rm m} \equiv \Omega_{\rm b0} + \Omega_{\rm D0}$ and $t_{\rm U}$ is the age of the universe, are presented in Fig.~\ref{fig:4}. First, considering the base+BAO+$H_0$ dataset alleviates the tension with SH0ES since for a $\Lambda$CDM the average Hubble constant is $\bar h=0.687$, whereas in the  $\Lambda \beta$CDM it is $\bar h=0.698$. The change in the $\chi^2$ at the posterior maximum, depending on the inclusion or not of $H_0$ in the previous dataset, noted $Q_{\rm DMAP} \equiv \sqrt{\Delta \chi^2}$ (see Refs.~\cite{Schoneberg:2021qvd,Khalife:2023qbu} for the methodology), is $Q_{\rm DMAP} =3.6$. On the other hand the AIC criterium of $\Lambda \beta$CDM relative to $\Lambda$CDM is $\Delta {\rm AIC} =-2.6$ which indicates only a marginal improvement. However, we need to highlight that the low-$z$  BAO are in tension with DES Y1~\cite{DES:2017myr}, favoring a rather high marginal value of matter fraction today ($\Omega_{\rm m} =0.388 \pm 0.050$) whereas the  galaxy clustering and weak lensing from DES Y1 constrain it to $\Omega_{\rm m}=0.248^{+0.030}_{-0.017}$, hence an approximate $2.5\sigma$ tension.\footnote{Note that DES Y3 results \cite{DES:2021wwk} are shifted upwards with $\Omega_{\rm m}=0.339^{+0.032}_{-0.031}$.} High-$z$ BAO are, however, independently very consistent with a lower $\Omega_{\rm m} = 0.254\pm0.030$; see also Fig.~5 of Ref.~\cite{eBOSS:2020yzd}. Besides, the low-$z$  BAO are also the most sensitive to the fiducial cosmology used in the analysis. This motivates us to perform an analysis with the base+BAO($z>1$)+$H_0$ dataset; see Ref.~\cite{PU_2} for a full argumentation of this choice. This leads to $\bar h=0.719$, in very good agreement with SH0ES, while other criteria also improve substantially ($Q_{\rm DMAP} =2.0$ and $\Delta{\rm AIC} =-14.5$). As anticipated the marginal matter fraction, $\Omega_{\rm m}=0.2666\pm 0.0073$, is much lower than the $\Lambda$CDM constraint with the same dataset ($\Omega_{\rm m}=0.2912\pm 0.0052$) so as to maintain the same $R_{\rm ang}$ with a larger $H_0$. As a consequence, in the $\Lambda\beta$CDM model, the universe is younger with $t_{\rm U} = 13.55\pm0.05\,{\rm Gy}$, an age consistent with the value deduced from globular clusters (GC) \cite{Valcin:2021jcg,Bernal:2021yli}. Note also that $S_8 = 0.807\pm 0.010$, hence the tension with DES results increases only mildly. 

Since our best fit model preserves the sound horizon and the physical content before the matter/radiation equality, the residuals with the CMB data are nearly as good as for the $\Lambda$CDM best fit; see Fig.~\ref{fig:3}. The reduction of DM density of order 5\% due to the nonminimal coupling implies that in the matter era the universe expands slower than the $\Lambda$CDM that would begin in the same conditions, but eventually it expands faster once the larger cosmological constant dominates, leading to the same comoving distance to the CMB surface, but with a larger $H_0$. This famous degeneracy line in the $(\Omega_{\rm m},H_0$) plane is obvious in Fig.~\ref{fig:4}, but the relative heights of the acoustic peaks, which require a fixed ratio between DM, photons and baryons around recombination, select only a region in it. By triggering the disappearance of DM, the $\Lambda\beta$CDM model selects another region of this degeneracy line with a lower $\Omega_{\rm m}$ hence allowing for a larger Hubble constant. The small tension on $\Omega_{\rm b}h^2$ mentioned in Refs.~\cite {Pitrou:2020etk,Pitrou:2021vqr} remains of the same order, since this model is precisely built to avoid any alteration of BBN physics.

\vskip0.25cm
 \noindent{\bf Discussion.} The $\Lambda\beta$CDM model is a simple and minimal extension of the $\Lambda$CDM with only one extra parameter in which the physics of the SM sector remains fully unchanged. It assumes that DM experiences ST gravity. Since $\rho_\varphi$ is subdominant during the whole cosmic history, it implies that (1) BBN predictions remain fully unaffected; (2) the low-$z$ expansion rate is unchanged compared to the standard $\Lambda$CDM; and (3) it escapes {\em by construction} all existing local constraints on the deviation from GR in the Solar system and in particular on the variation of $G$ since, again, SM fields are transparent to the scalar interaction; there is no testable violation of the WEP~\cite{MICROSCOPE:2022doy} or variation of a constant~\cite{Uzan:2002vq}.  

We explored a minimal model in a fully consistent theory, assuming a massless scalar field and a quadratic coupling. This allowed us to compute unambiguously cosmological predictions both at the background and perturbation levels, going beyond many phenomenological parameterisations of interacting DM~\cite{DiValentino:2021izs,Schoneberg:2021qvd}. Note that the consistency of the theory, to which one shall attribute a credence compared to {\it ad hoc} or not fully predictive constructions, is not taken into account in model comparisons~\cite{DiValentino:2021izs,Schoneberg:2021qvd}, as well as  noncosmological constraints, which our model avoids. The MCMC analysis of the latest data shows that the $H_0$ tension reduces to $3.8\sigma$ and reaches below $2\sigma$ when low-$z$ BAO data  are discarded while the $\Lambda$CDM remains marginally improved from $4.4\sigma$ to $4.1\sigma$ (see Table~\ref{tab1}). This confirms the insight of Ref.~\cite{Vagnozzi:2023nrq} that ``new physics is not sufficient to solve the $H_0$ problem.''

Indeed, several models have already considered a coupling of DM to a scalar field. Refs.~\cite{Damour:1990tw,Coc:2008yu,Fuzfa:2007sv,Agrawal:2019dlm} argued for such a coupling from the swampland conjecture to alleviate the Hubble tension while Ref.~\cite{Gomez-Valent:2023jgn} showed that a coupling to DE is not able to alleviate the $H_0$ tension. In the model by Ref.~\cite{Thomas:2022ucg} the transition occurs at too low a redshift. Several models~\cite{Liu:2023kce,vandeBruck:2022xbk,Bottaro:2023wkd,Archidiacono:2022iuu} were also built with scalar DM coupled to DE. We recall that a key ingredient of $\Lambda\beta$CDM is that $\varphi$ never dominates  and eventually decays faster than radiation, so that it is not a DE model and provides an efficient mechanism to extract part of DM, while marginally  imprinting the matter spectrum. Generically, the model allows the cosmology to have the same sound horizon as a $\Lambda$CDM with a higher $H_0$ and a lower $\Omega_{\rm D0}$. The extra parameter $\beta$ controls when the transition occurs with respect to equality, and $A(\varphi_i)$ determines the magnitude of this effect.  It can be extended in many ways leading to a whole family of models described  and investigated in our companion article~\cite{PU_2}. The physics of the dark sector occurs naturally around the LSS due to the coupling to DM so that it does not fall in the ``late/early" distinction~\cite{DiValentino:2021izs,Abdalla:2022yfr} as suggested in Ref.~\cite{Ye:2020btb}. Interpreted as a varying $G$ model~\cite{Begue:2017lcw,Knox:2019rjx}, it avoids {\em by construction} the difficulties with either BBN and/or local constraints of GR. While our $\Lambda\beta$CDM model is almost indistinguishable from the
$\Lambda$CDM, it enjoys the specific feature that DM and baryons do not feel the same gravity that will imprint their velocity, and hence modify
the measurement of velocity induced acoustic oscillations~\cite{Tseliakhovich:2010bj} that could be measured from 21cm observations~\cite{Munoz:2019rhi}, offering a unique window to probe our new proposal and more generally to test the equivalence principle between the visible and dark sectors for DM models.

We have to stress that the microphysics of DM has not been discussed. A dark fifth force may lead to a nonvanishing effects in E\"otv\"os tests that could be probed~\cite{Carroll:2009dw,Mantry:2009ay} in particular if DM interacts with SM fields. Such models are strongly constrained~\cite{Carroll:2008ub} and it was suggested that the anomalies in the positron/electron spectra may arise from a  dark force mediating the DM annihilation, possibly detectable at the LHC~\cite{Bai:2009it}. As a conclusion, this encouraging new model gives a  simple and minimal extension of the $\Lambda$CDM that is in good agreement with all cosmological data when SH0ES is not taken into account and that alleviates the Hubble tension with SH0ES and H0lLiCow~\cite{Wong:2019kwg} while being compatible with all local experiments and BBN.  It alleviates the $H_0$ tension to $3.8\sigma$ and resolves it to less than $2\sigma$ if we discard low-$z$- BAO data.  As a fully consistent theory it goes beyond any phenomenological parameterization, thus offering the possibility to be tested in any environment, e.g. DM halos~\cite{Frieman:1991zxc,Gradwohl:1992ue,Nusser:2004qu,Kesden:2006vz,Bean:2008ac}. The study and constraints on the gravitation of the dark sector can lead  to a better understanding of DM~\cite{Moody:1984ba,Bovy:2008gh}.

\bibliographystyle{ieeetr}
\bibliography{BIBHO}
\end{document}